# Terahertz optical activity near crystal field transitions of $Tm^{3+}$ ions in magnetoelectric alumoborates

A.M. Kuzmenko [1], V.Yu. Ivanov [1], S.V. Garnov [1], A. Shuvaev [2], A. Pimenov [2], K.N. Boldyrev [3], I. A. Gudim [4], and A. A. Mukhin [1*]

[1] Prokhorov General Physics Institute of the Russian Academy of Sciences, 119991 Moscow, Russia
[2] Institute of Solid State Physics, Vienna University of Technology, 1040 Vienna, Austria
[3] Institute of Spectroscopy of the Russian Academy of Sciences, 142190 Moscow, Russia
[4] Kirensky Institute of Physics, Siberian Branch of the Russian Academy of Sciences, 660036 Krasnoyarsk, Russia

**Abstract**
Crystal field (CF) excitations in the ground multiplet $^3H_6$ of $Tm^{3+}$ ions were investigated using terahertz transmission spectra of magnetoelectric $TmAl_3(BO_3)_4$ and $Tm_{0.05}Yb_{0.1}Y_{0.85}Al_3(BO_3)_4$. These excitations were identified as mainly magnetic dipole transitions from the ground singlet $A_1$ to the next excited doublet E, split by the crystal field of the $D_3$ symmetry. The fine structure of the modes was resolved at low temperatures. It manifested differently in lightly doped and in pure Tm borates, consistent with different distortions of the local crystal field with the $D_3$ symmetry. Strong natural optical activity was observed near the CF transitions resulting in a polarization plane rotation up to 25 degrees. The optical activity is quantitatively described by contributions of magnetic and electric dipole transitions to dynamic magnetoelectric susceptibility and taking into account the classification of local distortions.

## I. Introduction

Rare-earth alumoborates $RAl_3(BO_3)_4$ (and ferroborates $RFe_3(BO_3)_4$) are new families of magnetoelectrics (multiferroics) with noncentrosymmetric trigonal crystal structure. They attracted a considerable attention due to interesting magnetoelectric (ME), magnetic, and optical properties which depend strongly on the rare-earth (R) ions and their crystal field (CF) states [1-5]. These compounds have trigonal structure (space group R32) of natural mineral huntite [6] built of helicoidal chains of edge-sharing $AlO_6$ ($FeO_6$) octahedra along the trigonal c-axis, interconnected by $BO_3$ triangles and $RO_6$ prisms (Fig. 1). The $R^{3+}$ ions occupy the $D_3$ symmetry positions inside the isolated prisms with no direct R-O-R bonds. The rare-earth ions play a key role in magnetoelectric properties of borates, with especial emphasis on their low energy CF states which are usually located in the terahertz frequency range. Both magnetic and electric dipole transitions between the CF states are responsible for the magnetoelectric effect. According to Ref. [7], an electric polarization in rare earth borates is determined by electric dipole moment of the rare-earth $4f$-shell as well as by the ion displacement induced by external magnetic or R-Fe exchange fields (single ion mechanism). This allowed to describe the temperature and field dependences of electric polarization in Nd, Sm, and Eu ferroborates [7] as well as in alumoborates with $R$ = Tm, Ho, Tb [4, 8].

Dynamic magnetoelectric effects appear via spectroscopic signatures in electromagnetic spectra in the vicinity of resonant excitations and they were studied in various magnetoelectric materials at optical frequencies [9-12], in the terahertz [13-15], and in the microwave [16-18] ranges. In ferroborates the giant optical activity and dichroism were observed in antiferromagnetic $SmFe_3(BO_3)_4$ [19-21] near the antiferromagnetic electromagnon mode [22] that can be excited either by magnetic or by electric components of light. Similar magnetoelectric polarization rotation was observed in paramagnetic alumoborate $YbAl_3(BO_3)_4$ near the transitions of the ground $Yb^{3+}$ doublet, split by magnetic field [23]. These studies were performed in external magnetic fields in order to increase the frequencies of the corresponding excitations to the sub-THz range. In such experiments the sign of the dynamical magnetoelectric effect changes after the field inversion.

---

[*] mukhin@ran.gpi.ru

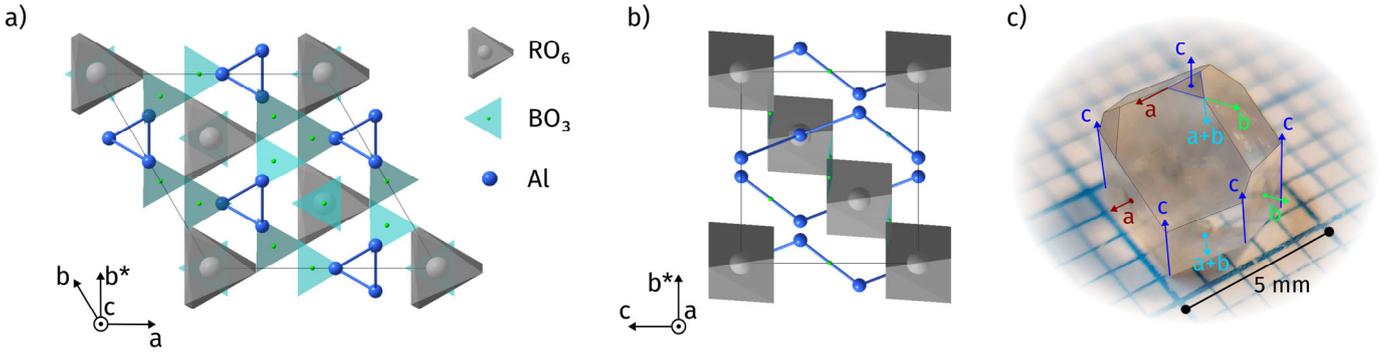

Fig. 1. Trigonal crystal structure (R32) of rare-earth aluminum borate projected onto the *ab*-plane (a) and *b\*c*-plane (b) (*b\** is perpendicular to the *a* and *c* axes). (c) The orientation of the crystallographic axes with respect to natural crystal faces in the source crystal used to prepare *a*- and *c*-cut samples.

The present work is devoted to terahertz study of a *natural* optical activity at low-energy $Tm^{3+}$ crystal field transitions in $TmAl_3(BO_3)_4$ borates and in Tm-doped $YAl_3(BO_3)_4$ alumoborates in zero magnetic field. The electronic structure of $Tm^{3+}$ in lightly doped $YAl_3(BO_3)_4$ crystals has been determined recently [24] by absorption and emission spectroscopy in the 5000–30000 $cm^{-1}$ range. It was found that the ground crystal-field state of the $Tm^{3+}$ ions is represented by a singlet ($A_1$) and the next excited state is a doublet (E) at $\nu\sim29$ $cm^{-1}$, while the remaining crystal field states are above 100 $cm^{-1}$. Similar structure of the low-energy crystal field states of $Tm^{3+}$ in undoped $TmAl_3(BO_3)_4$ was confirmed by direct analysis of A↔E transitions of the ground $Tm^{3+}$ multiplet. The fine structrure due to local distortions was observed as well [25]. Here we present a detailed study of the CF transitions and their fine structure by terahertz transmission and by polarization rotation experiments. Theoretical calculations allowed to extract contributions of magnetic- and electric-dipole transitions responsible for the optical activity and to clarify specific features of the fine structure in pure and diluted Tm-borates.

## II. Experiment

Single crystals of $TmAl_3(BO_3)_4$ and $Tm_{0.05}Yb_{0.1}Y_{0.85}Al_3(BO_3)_4$ were grown from a $Bi_2Mo_3O_{12}$–$Li_2MoO_4$–$B_2O_3$ based flux as described in Ref. [26]. The flux composition was 85 wt.% ($Bi_2Mo_3O_{12}$ + 0.5 $Li_2MoO_4$ + 2 $B_2O_3$) and 15 wt.% of the corresponding $RAl_3(BO_3)_4$ compound. The crystals were grown on seeds from the solution at temperatures around 1000–1050 °C with a rate below 0.5 mm/day, which is known to yield high-quality $RAl_3(BO_3)_4$ crystals with narrow optical transitions [26-29]. Despite a small partial substitution of rare earth by Bi ions from the flux [25], the crystal symmetry (space group R32) is preserved [30] and a good natural faceting is also observed, which allows to determine the crystallographic directions (Fig.1c). Oriented samples were prepared in the form of plane-parallel *c*-cut and *a*-cut plates with several thicknesses from 0.18 up to 3.5 mm and lateral dimensions up to 10 mm. Measurements of the polarized transmission spectra $T(\nu)$ were carried out using quasioptical backward-wave-oscillator (BWO) technique (Fig.2) [31] in the frequency range 10 – 39 $cm^{-1}$ (0.3 – 1.17 THz) and at temperatures 4 - 300 K. BWOs are tunable sources of electromagnetic radiation that rely on periodic modulation of an electron beam within a high-vacuum tube. The variation of the output frequency is realized by changing the accelerating voltage in the range of few kilovolts. The output power of the BWO's is typically 1-100 mW. An opto-acoustic Golay cell or a He-cooled bolometer were used as detectors of the radiation.

Transmission experiments on samples with different thickness are necessary to increase the dynamic range of the spectrometer especially in the case of strong absorption. To extract the mode parameters, the spectra of plane-parallel plates with different thickness were fitted self-consistently.

Strong resonance absorption was observed around $\nu\sim29$ $cm^{-1}$ both in $TmAl_3(BO_3)_4$ (Fig. 3) and $Tm_{0.05}Yb_{0.1}Y_{0.85}Al_3(BO_3)_4$ (Fig. 4) for the polarization with magnetic component ***h*** of light perpendicular to the trigonal *c*-axis and independently with the electric component ***e***. Periodic oscillations in transmission (Fig.4, left panels) are due to Fabry-Pérot interferences in a thick sample (d=3.547 mm) used for diluted system.

To determine the optical activity, the measurements were performed for the analyzer oriented parallel to the polarizer and rotated by $\alpha_A$ = +45 and -45 degrees, respectively. The observed difference between the transmission spectra for $\alpha_A$ =±45° clearly indicates the existence of the optical activity. To extract the polarization plane rotation $\Theta$ and the ellipticity $\eta$ we used the relations [19, 21]:

$$\tan 2\Theta = \tfrac{1}{2}(T_{45} - T_{-45})/(T_0 - T_{90}),$$
$$\tan^2 \eta = (E_{min}/E_{max})^2 = [1 - \tfrac{1}{16}(T_{45} - T_{-45})^2 / T_0 T_{90}]^{1/2}$$
(1)

where $T_0$, $T_{45}$, $T_{-45}$, $T_{90}$ are transmission coefficients for $\alpha_A$ = 0, ±45°, and 90°, respectively. Only three of four spectra are independent here. In particular, $T_{90}$ can be calculated from $T_0$, $T_{45}$, $T_{-45}$ using $T_{90} = (T_{45} - T_{-45} - 2T_0)/2$.

The incident radiation was linearly polarized along different crystallographic axes. This set of measurements provides information about the polarization plane rotation and the ellipticity. The same method was also used in Refs. [19, 21].

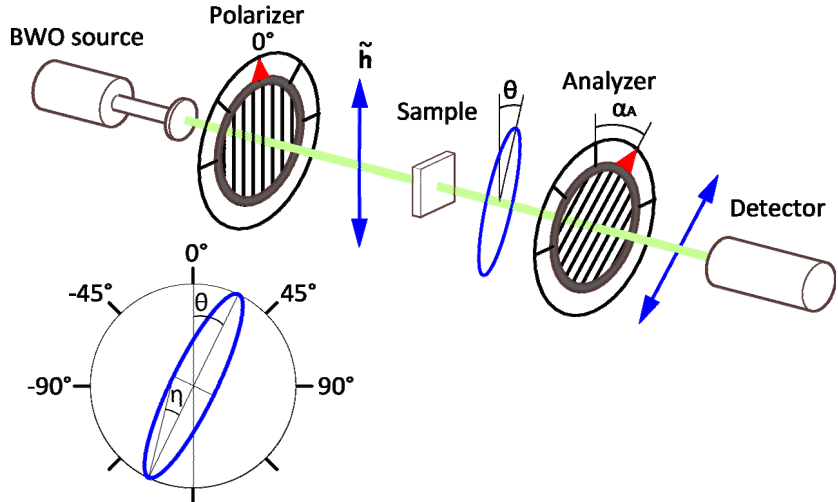

Fig. 2. Scheme of the optical activity measurements using the BWO spectrometer. Blue arrows indicate an alternating magnetic field for a linear polarization of the incident radiation. Blue ellipse – elliptical polarization of the radiation after transmission through the sample. The inset shows the definitions of the ellipticity $\eta$ and polarization rotation $\theta$.

## III. Results and discussion

The quasioptical terahertz spectra in pure TmAl$_3$(BO$_3$)$_4$ and diluted YAl$_3$(BO$_3$)$_4$:5%Tm$^{3+}$ (Figs. 3, 4) clearly demonstrate strong resonances around 25-33 cm$^{-1}$. With decreasing temperature their intensity increases as well as a fine structure is observed. These modes can be identified as CF transitions from the ground singlet A$_1$ to the next excited doublet E of the ground multiplet $^3H_6$ of Tm$^{3+}$ ions similar to Ref. [25]. The fact that they are observed only in the polarization of *ac* magnetic field **h** perpendicular to the c-axis, confirms their predominantly magnetic-dipole character.

The observed fine structure was assigned to local distortions of the sites with the D$_3$ symmetry by internal strains or by Bi$^{3+}$ impurities. This resulted in splitting of the A$_1$→E transition for two components E$^\pm$ of the doublet [25]. At least two types of locally distorted sites were observed in pure system [25] while only one kind of such sites could be detected in the diluted samples (Fig. 4).

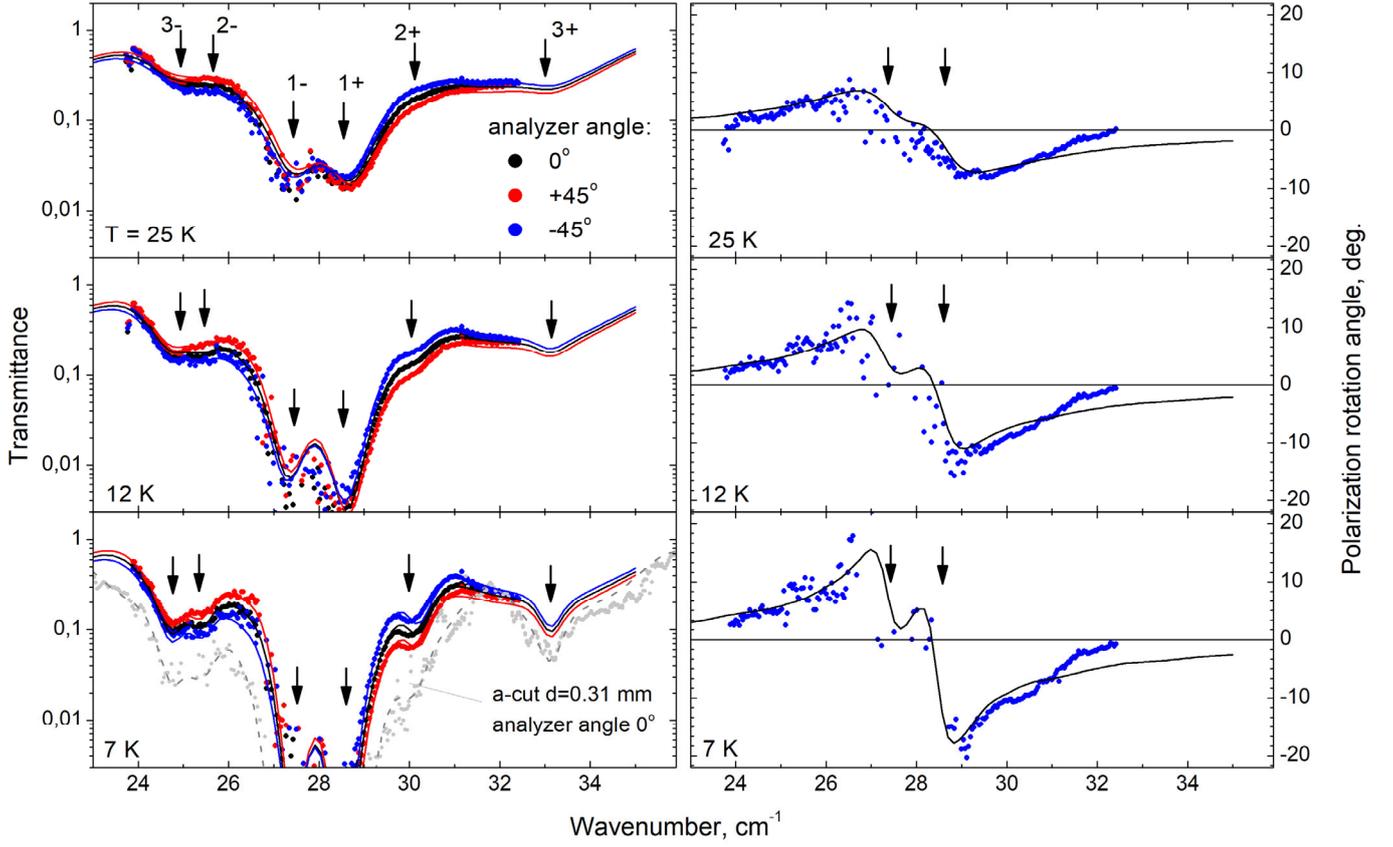

Fig. 3. Transmission spectra (left) and polarization rotation angle (right) of the c-cut TmAl$_3$(BO$_3$)$_4$ sample of thickness d=0.179 mm for *ac* magnetic field ***h***⊥*c*-axis. The spectrum for a-cut (grey) is shown to demonstrate additional resonance modes in the spectrum for $\nu >$ 32cm$^{-1}$. Points – experiment, solid lines – theory as described in the text.

We attempted to describe the pure TmAl$_3$(BO$_3$)$_4$ spectra by a model based on random lattice deformations similar to that used for the diluted crystal (see also Refs. [32–34]). However, such a model produces a single inhomogeneously broadened line and cannot reproduce the well-resolved three-doublet structure and the relative intensities of the six components observed simultaneously in the transmission and in optical-activity spectra. In contrast, the cluster model with three types of Tm$^{3+}$ sites in the vicinity of a Bi$^{3+}$ impurity on the R site, inspired by the multiplicities of the nearest-neighbor shells in the *R*32 structure [27], yields a unique set of parameters that fits both the transmission and the polarization-rotation spectra. Therefore, we attribute the fine structure in the pure Tm compound predominantly to local distortions produced by Bi$^{3+}$ impurities, whereas random internal strains give only a minor additional broadening.

A remarkable feature of the observed CF transitions is the natural optical activity, i.e., the rotation of the polarization plane. The corresponding spectra were determined applying Eq. (1) to the transmission spectra $T_0$, $T_{45}$, $T_{-45}$ (Fig. 3, 4, right). The maximum value of the rotation angle about 20-25° was observed in both pure (thin) and diluted (thick) samples.

The optical activity in the studied noncentrosymmetric crystals (R32) is allowed by symmetry and it can be explained by contributions of the magnetic-dipole and electric-dipole transitions to the dynamic magnetoelectric susceptibility $\kappa_{xx,yy}(\omega)$. The ME susceptibility determines the circular birefringence and dichroism and it transforms the incident linearly-polarized wave to an elliptical one (see, for example, Ref. [35]). Because the spectra in pure TmAl$_3$(BO$_3$)$_4$ and in diluted borates are substantially different, we consider them separately.

Magnetic, dielectric and magnetoelectric permittivities for **pure TmAl$_3$(BO$_3$)$_4$** include contributions from the observed transitions at several Tm$^{3+}$ sites (Fig. 3). Compared to Ref. [25], a thinner sample was available in our experiments. This allowed to resolve the three pairs of components (6 modes) for three types of distorted Tm positions. Sites #3 occupy six positions nearest to Bi impurity and they experience strong distortions that results in a largest splitting of the E doublet (Fig. 5). Sites #2 occupy six next-nearest positions. Finally, sites #1 are attributed to the remaining positions at larger distances and they have only a minor splitting of the E doublet. For small concentrations of Bi impurities $x_{Bi}$ the fractions of different sites are given by: $c_{2,3}= 6x_{Bi}$ for the sites #2, #3 and $c_1=1-12x_{Bi}$ for the sites #3, respectively. The energy levels for all three sites are depicted in Fig. 6.

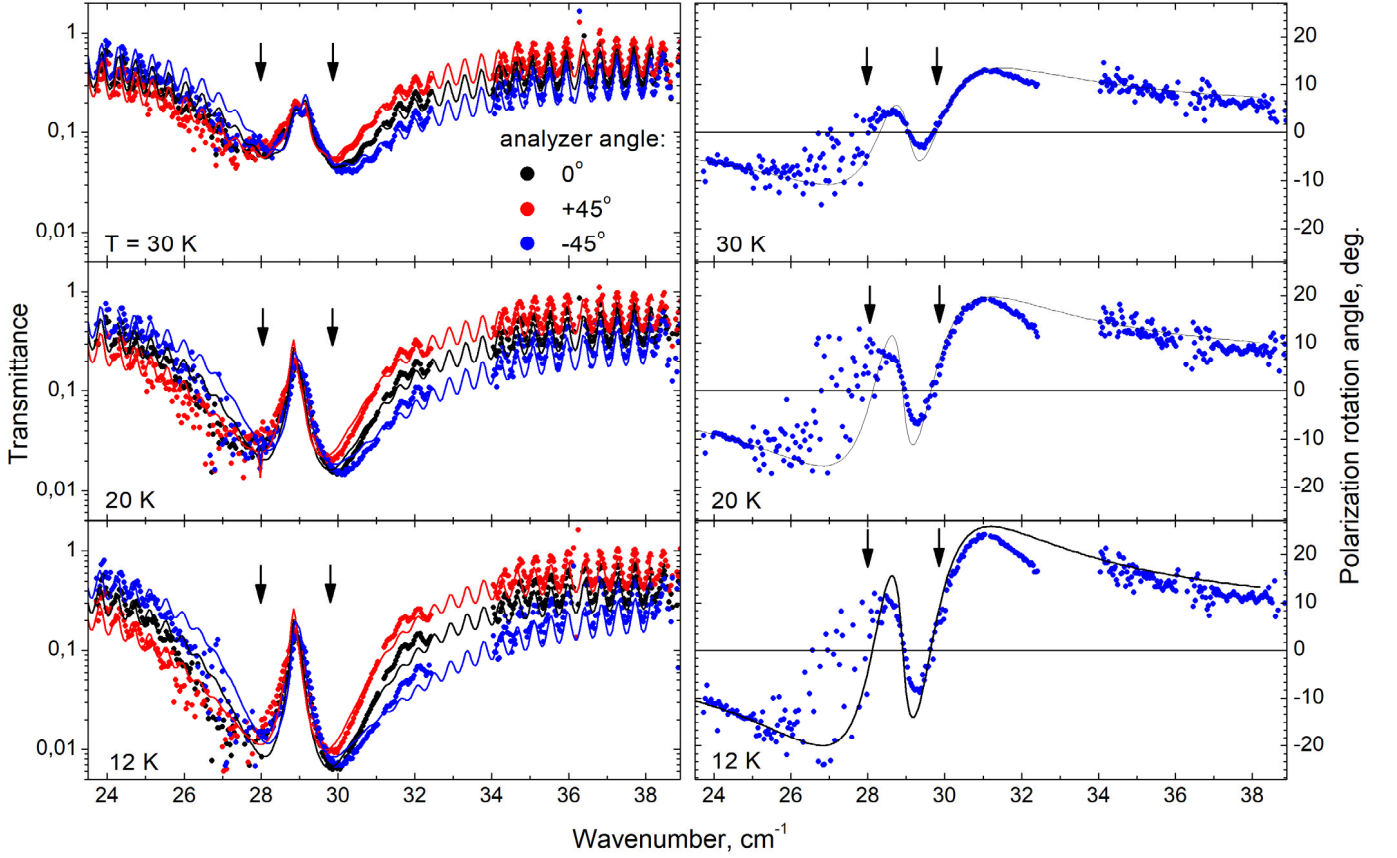

Fig. 4. Transmission spectra (left) and polarization rotation angle (right) of the c-cut Tm$_{0.05}$Yb$_{0.1}$Y$_{0.85}$Al$_3$(BO$_3$)$_4$ sample (d=3.547 mm) for *ac* magnetic field **h**⊥*c*-axis. Points – experiment, solid lines – theory as described in the text.

For the present geometry with **k**||*c*-axis (c-cut) the relevant components of magnetic, dielectric and magnetoelectric permittivities, $\mu_{xx,yy}$, $\varepsilon_{xx,yy}$, $\kappa_{xx,yy}$ can be represented as superpositions of the corresponding transitions

$$\mu_{xx,yy}(\omega) = 1 + 4\pi N \cdot \sum_{k,\pm}\left[c_k \cdot \frac{2(\mu_k^\pm)^2}{Z_k(E_k^\pm - E_0)}\left(e^{-\frac{E_0}{k_BT}} - e^{-\frac{E_k^\pm}{k_BT}}\right) \cdot R_L(\omega,\omega_k^\pm)\right],$$

$$\varepsilon_{xx,yy}(\omega) = \varepsilon_\perp + 4\pi N \cdot \sum_{k,\pm}\left[c_k \cdot \frac{2(d_k^\pm)^2}{Z_k(E_k^\pm - E_0)}\left(e^{-\frac{E_0}{k_BT}} - e^{-\frac{E_k^\pm}{k_BT}}\right) \cdot R_L(\omega,\omega_k^\pm)\right], \quad (2)$$

$$\kappa_{xx,yy}(\omega) = i \cdot 4\pi N \cdot \sum_{k,\pm}\left[c_k \cdot \frac{2(d_k^\pm \mu_k^\pm)}{Z_k(E_k^\pm - E_0)}\left(e^{-\frac{E_0}{k_BT}} - e^{-\frac{E_k^\pm}{k_BT}}\right) \cdot \frac{\omega}{\omega_k^\pm}R_L(\omega,\omega_k^\pm)\right],$$

where $E_0$ and $E_k^\pm = \hbar\omega_k^\pm$ (k=1,2,3) are the CF states of three types of sites of the ground singlet A$_1$ and of two components of the split E doublet, respectively; $d_k^\pm, \mu_k^\pm$ are effective matrix elements of the electric and magnetic dipolar transitions averaged over six equivalent Tm$^{3+}$ positions; $c_k$ is the concentration of the $Tm^{3+}$ ions in distorted sites of $k$-type (Fig. 6); $R_L(\omega, \omega_k^{\pm}) = \omega_k^{\pm 2}/(\omega_k^{\pm 2} - \omega^2 + i\gamma_k^{\pm}\omega)$ is the Lorenzian function describing the line shape of the $E_0 \to E_k^{\pm}$ transitions; $Z_k = \exp(-E_0/k_BT) + \exp(-E_k^+/k_BT) + \exp(-E_k^-/k_BT) + ...$ is a partition function, $N$ is the number of $Tm^{3+}$ ions per cm$^3$; $\varepsilon_{\perp}$ is the dielectric permittivity of the crystal in the basal plane.

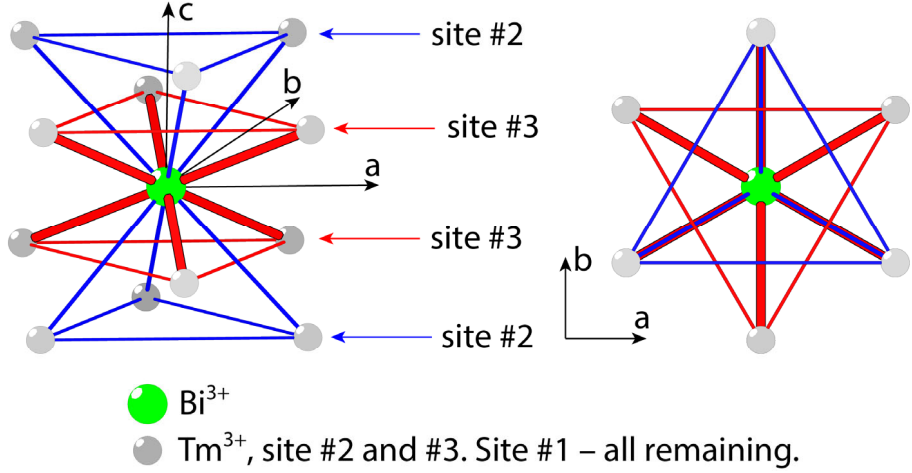

Fig. 5. Local arrangement of six possible positions of strongly distorted $Tm^{3+}$ ions around a single Bi-impurity for nearest neighbors (site #3) and next nearest neighbors (site #2). All remaining $Tm^{3+}$ ions occupy sites #1, with larger Bi–Tm separations than at sites #2 and #3 (not shown).

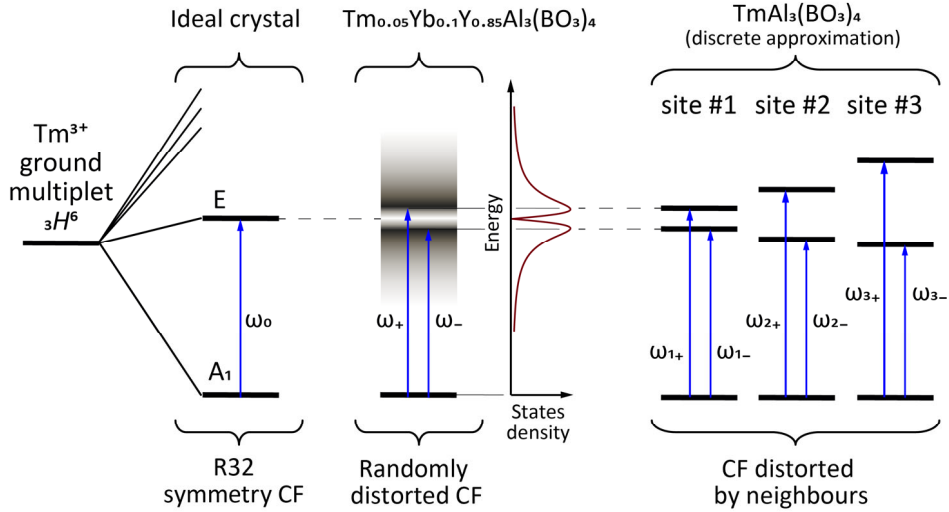

Fig. 6. Splitting of the crystal field states of $Tm^{3+}$ ions due to distortions in the vicinity of a Bi-impurity for different sites in pure $TmAl_3(BO_3)_4$ (right) and simulated continuous distribution of the E-doublet splitting in diluted $Tm_{0.05}Yb_{0.1}Y_{0.85}Al_3(BO_3)_4$ (middle).

The violation of the local symmetry due to distortions near $Bi^{3+}$ impurities leads to the splitting of the E-doublet and to changes in the wave functions and in selection rules for the matrix elements of transitions. These changes modify the symmetry of local magnetic, magnetoelectric, and dielectric susceptibilities. However, after averaging over the six $Tm^{3+}$ nearest neighbors (or next nearest neighbors) around a $Bi^{3+}$ impurity, additional contributions to the susceptibilities are canceled. Only contributions allowed by the original trigonal symmetry of the crystal remain. Therefore, in the first approximation we take into account only the effective (due to averaging) matrix elements allowed by the ($D_3$) trigonal symmetry. In this approximation, we take into account the violations of the local D₃ symmetry by introducing the doublet splitting E±, as well as by varying the values of magnetic and electric matrix elements. In this approximation the effects of lower local symmetry on the susceptibilities can be neglected due to averaging.

Using Eqs. (2), we simulated the transmission and optical activity spectra taking into account such features like Fabry-Pérot interferences [19]. A reasonable agreement with experiment was obtained including the temperature behavior as well. As a result, the relevant parameters of the CF transitions (energies, effective magnetic dipole and electric dipole moments) could be obtained (see Table 1). Electric dipole moments for the sites #2 and #3 could not be resolved due to small concentrations of these sites and due to weak intensities of the resonance modes. We obtained the Bi impurities concentration as $x_{Bi}$ = 0.022 ± 0.002 giving $c_{2,3}$ ~0.13 which is about 13% of $Tm^{3+}$ ions belonging to the two most distorted sites #2 and #3. This fraction is rather small compared to the contribution of nearly distorted sites #1 and is physically reasonable in view of independent spectroscopic estimates of Bi incorporation in similar crystals. High-resolution optical spectroscopy of $Yb^{3+}$ ions in stoichiometric $YbAl_3(BO_3)_4$ and in $Y_{1-x}Yb_xAl_3(BO_3)_4$ crystals grown from the same $Bi_2Mo_3O_{12}$–$Li_2MoO_4$ based flux has revealed several types of nonequivalent $Yb^{3+}$ centers associated with bismuth and molybdenum impurities occupying the rare-earth and aluminium sites [27-29]. The fraction of distorted $Yb^{3+}$ sites ($Yb^{3+}(Bi^{3+})$ and $Yb^{3+}(Mo^{3+})$ centers) was found to reach ≲1.4% in stoichiometric $YbAl_3(BO_3)_4$ [27] and it increases in $Y_{1-x}Yb_xAl_3(BO_3)_4$ crystals and in samples grown with deliberately modified flux composition [28, 29]. Even larger $Bi^{3+}$ contents, up to 5–7% on the rare-earth site, were independently deduced in iron borates grown from Bi-containing fluxes of similar type [36]. These results demonstrate that an uncontrolled incorporation of several atomic percent of $Bi^{3+}$ into the rare-earth sublattice is a generic feature of $RAl_3(BO_3)_4$ and $RFe_3(BO_3)_4$ crystals grown from $Bi_2Mo_3O_{12}$ -based fluxes. In this context, the value $x_{Bi} \simeq 2\%$ extracted from our fit of the $TmAl_3(BO_3)_4$ spectra is fully consistent with the known range of Bi concentrations in related materials and provides additional support for the Bi-impurity model.

The same parameters were used in the simulation of the spectra at different temperatures except for the linewidth that increased from ~ 0.8 cm⁻¹ at 7 K to ~ 1.5 cm⁻¹ at 25 K. In general, the energies of the transitions and magnetic dipole moments agree well to that in Ref. [25] where another measurement geometry was used not allowing to observe the optical activity. We note also that the optical activity is inherent to $A_1 \rightarrow E$ transitions and it will be preserved in undistorted crystals as well.

Table 1. Parameters of the observed CF transitions of $Tm^{3+}$ ions (resonance frequencies, matrix elements of effective magnetic dipole and electric dipole moments) in both systems extracted from transmittance and optical activity spectra. The dielectric constant $\varepsilon_\perp$ =10.15 ± 0.05 was equal for both compounds.

| Compound | $Tm_{0.05}Yb_{0.1}Y_{0.85}Al_3(BO_3)_4$ | $TmAl_3(BO_3)_4$ | | | | | |
|---|---|---|---|---|---|---|---|
| | | site #1 | | site #2 | | site #3 | |
| CF resonance frequency (cm⁻¹) | $\omega_0$ | $\hbar\omega_{1+}$ | $\hbar\omega_{1-}$ | $\hbar\omega_{2+}$ | $\hbar\omega_{2-}$ | $\hbar\omega_{3+}$ | $\hbar\omega_{3-}$ |
| | 28.9 ± 0.1 | 28.5 ± 0.1 | 27.3 ± 0.1 | 30.1 ± 0.1 | 25.4 ± 0.1 | 33.4 ± 0.1 | 24.7 ± 0.1 |
| Magnetic dipole moment ($\mu_B$) | $\mu_\perp$ | $\mu_{1+}$ | $\mu_{1-}$ | $\mu_{2+}$ | $\mu_{2-}$ | $\mu_{3+}$ | $\mu_{3-}$ |
| | 3.7 ± 0.1 | 3.6 ± 0.1 | 3.6 ± 0.1 | 2.9 ± 0.2 | 3.4 ± 0.2 | 3.0 ± 0.2 | 4.2 ± 0.2 |
| Electric dipole moment (10⁻³ D) | $d_\perp$ | $d_{1+}$ | $d_{1-}$ | $d_{2+}$ | $d_{2-}$ | $d_{3+}$ | $d_{3-}$ |
| | 4.5 ± 0.3 | 7 ± 1 | 5 ± 1 | – | – | – | – |

Now we discuss the **diluted system $YAl_3(BO_3)_4$:$Tm^{3+}$**. Here only one pair of the $A_1 \rightarrow E^\pm$ CF transitions with asymmetric line shape is observed in the transmission spectra (Fig.4). In contrast to the pure system, a probability of finding $Tm^{3+}$ near the Bi impurity is very low. Therefore, the splitting of the Tm doublet occurs mainly due to random local distortions and due to stress distribution in a real crystal. Similar splitting of the optical transitions between the ground singlet and the doublet of excited multiplet of $Tm^{3+}$ was observed in diluted $LiYF_4$:$Tm^{3+}$ and assigned to random lattice deformations [32]. The line shape of the observed transitions was simulated taking into account the distribution of lattice deformations. Here we used the approach developed in Refs. [32, 33] to simulate the observed transmission and the optical activity spectra.

The splitting of the excited $Tm^{3+}$ doublet is determined by random deformations of $e_1=u_{xx}-u_{yy}$ and $e_2=-2u_{xy}$ (as well as $e_3=u_{yz}$ and $e_4=-u_{xz}$) which transform according to two-dimensional representation of the $D_3$ point group. As a result, the deformation interaction violating the $D_3$ crystal field symmetry is created. In order to determine the electrodynamics response (i.e. $\mu_{xx,yy}$, $\varepsilon_{xx,yy}$, and $\kappa_{xx,yy}$), we averaged their local values using the two-dimensional distribution function $g(e_1,e_2)$ [32, 33] (see also Ref. [34])

$$g(e_1,e_2) = \frac{\Gamma}{2\pi(e_1^2+e_2^2+\Gamma^2)^{3/2}}, \qquad (3)$$

where $\Gamma$ is the distribution width. As a result, the permittivities can be expressed in the form:

$$\mu_{xx,yy}(\omega) = 1 + \Delta\mu_\perp^0 \cdot R(\omega,\omega_0,T),$$
$$\varepsilon_{xx,yy}(\omega) = \varepsilon_\perp + \Delta\varepsilon_\perp^0 \cdot R(\omega,\omega_0,T), \qquad (4)$$
$$\kappa_{xx,yy}(\omega) = i \cdot \sqrt{\Delta\mu_\perp^0 \Delta\varepsilon_\perp^0} \cdot R'(\omega,\omega_0,T),$$

where $\Delta\mu_\perp^0 = 8\pi N\mu_\perp^2/\hbar\omega_0$, $\Delta\varepsilon_\perp^0 = 8\pi Nd_\perp^2/\hbar\omega_0$ are contributions to magnetic and dielectric permittivities from $A_1 \to E^\pm$ transitions. The spectral functions

$$R(\omega,\omega_0,T) = \int_0^\infty \left[\left(\frac{1-\exp(-\hbar\omega_+/k_BT)}{Z(T)}\cdot\frac{\omega_0 R_L(\omega,\omega_+)}{\omega_+} + \frac{1-\exp(-\hbar\omega_-/k_BT)}{Z(T)}\cdot\frac{\omega_0 R_L(\omega,\omega_-)}{\omega_-}\right)\cdot\frac{u\widetilde{\Gamma}}{(u^2+\widetilde{\Gamma}^2)^{3/2}}du\right],$$

$$R'(\omega,\omega_0,T) = \frac{\omega}{\omega_0}\int_0^\infty \left[\left(\frac{1-\exp(-\hbar\omega_+/k_BT)}{Z(T)}\cdot\frac{\omega_0^2 R_L(\omega,\omega_+)}{\omega_+^2} + \frac{1-\exp(-\hbar\omega_-/k_BT)}{Z(T)}\cdot\frac{\omega_0^2 R_L(\omega,\omega_-)}{\omega_-^2}\right)\cdot\frac{u\widetilde{\Gamma}}{(u^2+\widetilde{\Gamma}^2)^{3/2}}du\right]$$

,
were determined by convolution of Lorenzian, describing the local responses with the distribution function, Eq. (3), where we introduce new variables $u \sim (e_1^2+e_2^2)^{1/2}$ and $\tan\varphi = e_2/e_1$ and integrate analytically on the angle $\varphi$; $\hbar\omega_\pm = \hbar\omega_0 \pm u$ are local frequencies of the CF transitions with a random splitting $2u$, $\omega_0$ is the frequency of the non distorted transition, $\widetilde{\Gamma}$ is the distribution width, $R_L(\omega,\omega_k) = \omega_\pm^2/(\omega_\pm^2-\omega^2+i\gamma_\pm\omega)$ is a Lorentzian, and $Z(T)$ is the partition function.

The simulations of the transmission and optical activity spectra show a good agreement with the experiment. The extracted value of the distribution width $\widetilde{\Gamma} = 1.15 \pm 0.05$ cm$^{-1}$ could serve as a measure of deformations in the diluted crystal. The local linewidth $\gamma_\pm$ increases from $\sim 0.1$ cm$^{-1}$ at 12 K to $\sim 0.25$ cm$^{-1}$ at 30 K. The values of magnetic and electric dipole moments of the transitions in the diluted system are of similar magnitude as in the pure TmAl$_3$(BO$_3$)$_4$ (see Table 1).

## IV. Conclusion

In summary, we have shown that natural terahertz optical activity at crystal-field transitions of $Tm^{3+}$ in TmAl$_3$(BO$_3$)$_4$ and YAl$_3$(BO$_3$)$_4$:Tm can reach rotation angles up to 20–25° in zero magnetic field. By combining transmission and polarization-rotation measurements with a microscopic model, we were able to separate the magnetic- and electric-dipole contributions to the dynamical magnetoelectric response and to quantify the electric-dipole matrix elements. Our analysis demonstrates that local structural distortions – in particular Bi-related impurity centers in the pure compound and random lattice deformations in the diluted crystal – play a key role in shaping both the fine structure of the crystal-field spectra and the magnitude of the optical activity. These findings complement earlier studies of field-induced magnetoelectric optical activity in rare-earth borates [19-21, 23] and establish terahertz natural gyrotropy at crystal-field transitions as a sensitive probe of the local symmetry in noncentrosymmetric rare-earth crystals.

**Acknowledgments**

This work was supported by the Russian Scientific Foundation (project # 25–79–30006.) and by the Austrian Science Fund (FWF) (Grants DOI 10.55776/I5539 and 10.55776/PAT7680623).